\documentclass[11pt] {article}      
\usepackage{amsmath,amssymb, amsthm, eucal,accents}  
\usepackage{graphicx}
\def\Ker{\hbox{\rm Ker\,}}
\def\inn{\subset}
\def\scirc{\circ}
\def\FF{\mathcal F}
\def\XX{\mathcal X}
\def\LL{\mathcal L}

\def\mP{\bar{P}}
\def\midx{{\Bigg|}}
\def\zzz{\longrightarrow}

\newcommand{\dbar}{d\hspace*{-0.08em}\bar{}\hspace*{0.1em}}

\newcommand{\jjj}{\,\vrule width6pt height.4pt depth0pt
        \vrule height6pt width.4pt depth0pt\,}    
\usepackage{times}
\usepackage[T1]{fontenc}     
\usepackage{enumerate}       
\usepackage{float}                 
\usepackage{mathtools}
\usepackage{tikz}
\usetikzlibrary{matrix, arrows, calc, fit}
\usetikzlibrary{arrows, decorations.markings}
\tikzstyle{AArrow} = [thick, decoration={markings,mark=at position 1 with {\arrow[semithick]{open triangle 60}}},%
   double distance=1.4pt, shorten >= 5.5pt,
   preaction = {decorate},
   postaction = {draw,line width=1.4pt, white,shorten >= 4.5pt}]
\tikzstyle{AArroww} = [semithick, white,line width=1.4pt, shorten >= 4.5pt]
\usepackage[strict]{changepage}
\usepackage{relsize}
\theoremstyle{plain}
\newtheorem{theorem}{Theorem}[section]            

\newtheorem{corollary}[theorem]{Corollary}	      %

\theoremstyle{definition}
    
\newtheorem{remark}[theorem]{Remark}	
\newtheorem{example}{Example}[section]

\numberwithin{theorem}{section}
\numberwithin{equation}{section}
\numberwithin{figure}{section}
\newcommand{\gaction}[2]{\genfrac{}{}{0.5pt}{}{#1}{#2}%
                        \!\lower2pt\hbox{\rotatebox[origin=c]{-90}{{$\looparrowright$}}}}

\newcommand{\dotfraction}[2]{\genfrac{}{}{0.5pt}{}{#1}{#2}%
                        \!\lower.5pt\hbox{{$\circ$}}}
\def\q{\partial}

\renewcommand\appendix{\par
  \setcounter{section}{0}
  \setcounter{subsection}{0}
  \setcounter{figure}{0}
  \setcounter{table}{0}
  \renewcommand\thesection{Appendix \Alph{section}}
  \renewcommand\thefigure{\Alph{section}\arabic{figure}}
  \renewcommand\thetable{\Alph{section}\arabic{table}}
}
\usepackage{titlesec}                                                           
\titlelabel{\thetitle.\ }                                                           
\titleformat*{\section}{\fontsize{14pt}{14pt} \bf}                   
\usepackage{fancyhdr}
\usepackage{sidecap}  
\usepackage[hang,small,sc]{caption}

\def\smalll{\scriptsize}

\begin{document}

\title{{\bf On geometry of phenomenological thermodynamics}\\
{\small This is  the preprint version of the article [10] under the same title} }%

\author{Jerzy Kocik                  
\\ \small Department of Mathematics
\\ \small Southern Illinois University, Carbondale, IL62901
\\ \small jkocik{@}siu.edu  }
\date{}

\maketitle

\begin{abstract}
\noindent
We present the formalism of phenomenological thermodynamics in terms of the  
even-dimensional symplectic geometry,  
and argue that it catches its geometric essence in a more profound and clearer way than the  
popular odd-dimensional contact structure description.
Among the advantages are a number of conceptual clarifications:
the geometric role of  internal energy (not made as an independent variable), the lattice of potentials, 
and the gauge interpretation of the theory.
\\[7pt]
{\bf Keywords:} Symplectic geometry, thermodynamics, lattice of thermodynamic  potentials, Ehresmann connection.
\\[7pt]
{\bf MSC:} 
80A05, 
53D05,  
53Z05.  
\end{abstract}


\section{Motivation}
\label{sec:motive}

Phenomenological thermodynamics of equilibrium (PTE) has become a
standard axiomatic theory in hands of Carath\'eodory \cite{Ca} and Gibbs \cite{Gi}.  
Formulated in the general way \cite{Ti,Cl}, it reveals a universal
structure which became rather incorporated into the latter statistical and quantum statistical mechanics than replaced by them.%

It has often been remarked that phenomenological thermodynamics is an example of a theory the formalism of which cannot be understood without the language of differential geometry.  
Paradoxically, the standard expositions often leave the geometrical content unclear, the ``differential calculus'' is formal, often carried out without specified underlying manifolds --- leaving the student of PTE rather perplexed.%
\footnote{Carath\'eodory assumes that each description of a given
system requires independent treatment and is based on a different
underlying space with different coordinates.
Most often, the underlying manifold of PTE in
the classical textbooks is that of $n+1$ extensive variables
(like $S$, $V$, $N$, and $U$).
}

Recently, phenomenological thermodynamics joined the program of
geometrization in physics.%
\footnote{For development of this view, see 
\cite{He,Mi,Mr1,Mr2, Ti,Ch}  and  citations therein.}
These approaches extended the configuration space of PTE by including $2n+1$ coordinates 
(like $S$, $V$, $N$, $T$, $p$, $\mu$, and $U$) and describe PTE as an odd-dimensional ``phase space'' with a contact structure.

\newpage

In this exposition, we propose an alternative description in which 
\textbf{symplectic structure} appears as the one that underlines PTE.%
\footnote{This exposition closely follows \cite{Ko}.}
The internal energy $U$ is not a basic coordinate but rather contributes---together with other thermodynamical potentials---to an
{\it algebraic lattice} of potentials.
In particular, one needs to distinguish between the ``universal internal energy'' $U$ and the internal energy of a system $u$.

\section{The phase space of PTE}
\label{sec:tdopt.2}

Here are the basic definitions constituting PTE.


\leftskip=.2in
\begin{enumerate}
\item[$\bullet$]
\textbf{Thermodynamical phase space} is a pair $(M,\alpha)$, where M is a $2n$-dimensional manifold and $\alpha \in \Lambda^1M$ is a differential l-form of the maximal rank. 
The form $\alpha$ will be called the Gibbs form.
Consequently, the phase space forms also a symplectic manifold, since the bi-form $\omega = : d\alpha$ is nondegenerate.
A \textbf{state} is a point in $M$.
\\

\begin{example}    \label{exm:tdopt.2.1}
The standard model of thermodynamics is based on a 6-dimensional linear space with variables:

\begin{center}
\begin{tabular}{lll}
$T$   --- temperature         && $S$  --- entropy \\
$P$   --- pressure            && $V$  --- volume \\
$\mu$ --- chemical potential  && $N$  --- number of molecules  \\[2pt]
\hline\\[-12pt]
$[$intensive variables$]$       && $[$intensive variables$]$
\end{tabular}
\end{center}
\end{example}
Much distress can be avoided if we denote the negative pressure simply as 
$$
\mP \  \equiv \  -P \,.
$$
\item[]
The (adjusted) differential {\bf Gibbs form} is
%
\begin{equation}    
\label{eq:gibbs}
       \alpha = S\,dT + V\,d\mP + N\,d\mu
\end{equation}
The {\bf symplectic form} on $M$ is
\begin{equation}    
\label{eq:symplectic}
d \alpha = dS\wedge dT +dV\wedge d\mP + dN\wedge d\mu
\end{equation}

\item[$\bullet$]
A \textbf{process} is a curve $c: \mathbb R \to M$.
The one-form $\alpha$ has a physical sense of the work $W[c]$ associated with carrying out the process $c$:
\begin{equation}    \label{eq:td2.1}
  W[c] = \int_c {\alpha}
\end{equation}
Process $c$ is \textbf{admissible} if $\dot c\in \Ker\alpha$, that is $\langle\alpha|\dot c\rangle$ (implying $\int_c \alpha = 0$).  

\item[$\bullet$]
A \textbf{theory} (or a \textbf{system}) is a submanifold $\Lambda\inn M$ of the phase space such that the Gibbs form vanishes when restricted to $\Lambda$:
\begin{equation}   
\label{eq:td2.2}
  \alpha\mid_\Lambda = 0
\end{equation}
If the embedding map is denoted
\begin{equation}    \label{eq:td2.3}
  \imath : \Lambda \to M
\end{equation}
then (\ref{eq:td2.2}) can be written simply as
\begin{equation}    \label{eq:td2.2'}
  \imath^*\alpha = 0
			\tag{5.2$'$}
\end{equation}
Equation \eqref{eq:td2.3} defining system $\Lambda$ is called the \textbf{equation of state}.

\item[]
Equation \eqref{eq:td2.2} is called the \textbf{Gibbs-Duhem relation} 
and means that $\alpha$ restricted to $\Lambda$ vanishes, i.e., that the directions tangent to $\Lambda$ lie in the kernel of $\alpha$, i.e. $T\Lambda\inn \Ker\alpha$.

\begin{figure}[t]
\[
  \includegraphics[width=4.5in]{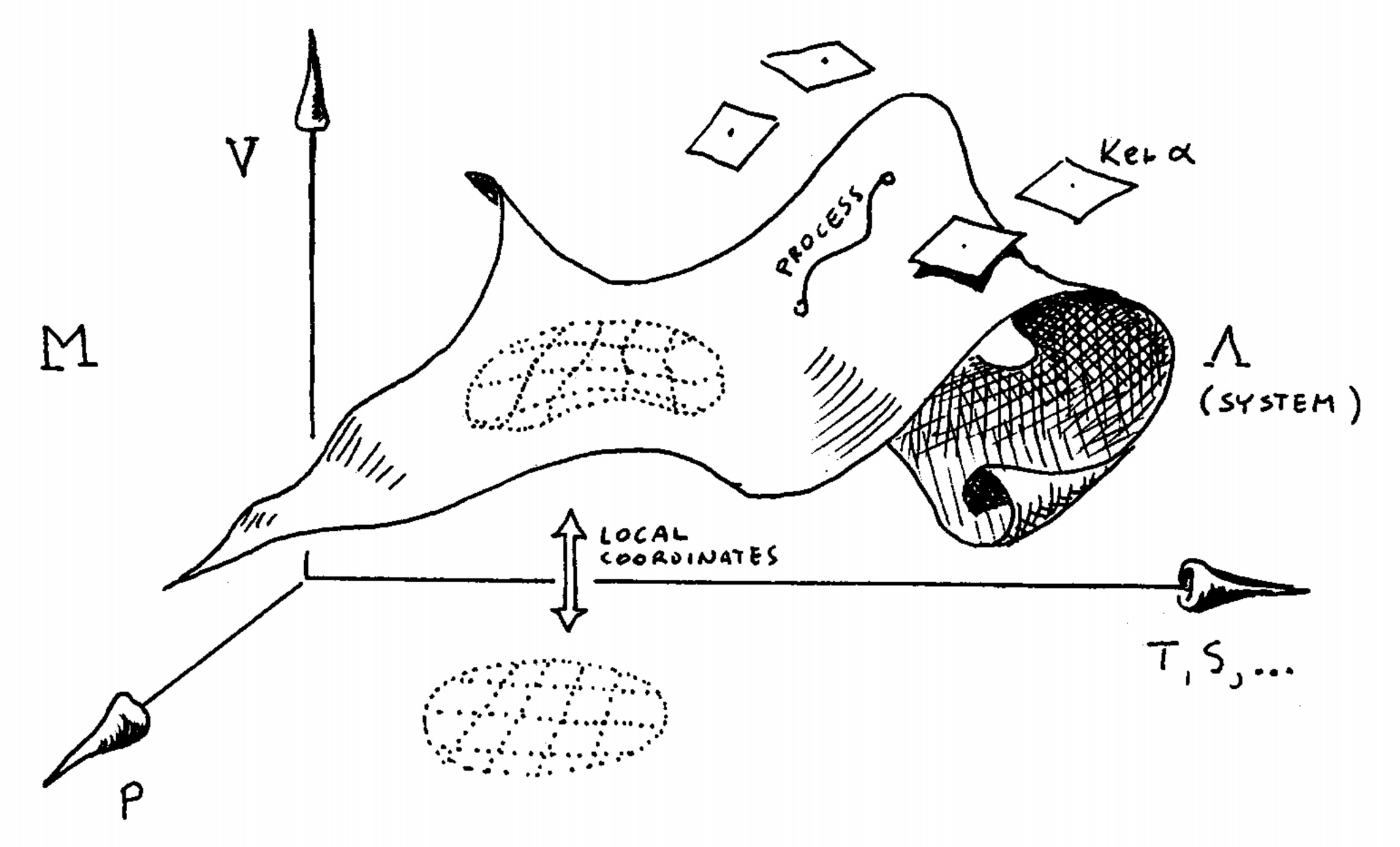}
\]
\caption{Phase space and a ``theory''}
\end{figure}

\begin{corollary}    \label{cor:td2.1}
Since the external derivative commutes with the map $\imath^*$, it immediately follows from \eqref{eq:td2.3} that 
\begin{equation}    \label{eq:td2.4}
  \imath^*d\alpha = 0,
\end{equation}
i.e., that $\Lambda$ is a Lagrangian submanifold of M (notice that \eqref{eq:td2.3} 
imposes conditions for $\Lambda$ more restrictive than \eqref{eq:td2.4}). 
This implies that the dimension of $\Lambda$ is at most ${1\over 2}\dim M$.
Equation \eqref{eq:td2.4} is called the \textbf{Maxwell identity} and is, in fact, 
the integrability condition for $\alpha$ on $\Lambda$
(it is an analog of Hamilton-Jacobi equation in classical mechanics).
\end{corollary}

\item[$\bullet$]
The \textbf{special phase space} is the cotangent bundle over a
\textbf{linear} space $Q$
%
%
%
\begin{equation} \label{eq:td2.5}
\begin{tikzpicture}[baseline=-0.8ex]
    \matrix (m) [ matrix of math nodes,
                         row sep=3em,
                         column sep=4em,
                         text height=4ex, text depth=3ex] 
 {
             \quad  M \quad =\quad  T^*Q \cong \mathbb R^{2n}  \phantom{\quad ~~~~~~~~~~}  \\
           \qquad\qquad\qquad   Q\cong \mathbb R^{n}   \phantom{\quad ~~~~~~~~~~}   \\
  };
    \path[-stealth]   (m-1-1) edge node[right] {$\pi$}  (m-2-1);
\end{tikzpicture}   
\end{equation}
The Gibbs form $\alpha$ is defined as the natural Liouville form
of the cotangent bundle.  
Manifold $Q$ will be called configuration space and coordinates on $Q$ are the \textbf{intensive parameters} of PTE.
If ${x'^i}$ are any linear coordinates on $Q$, and ${p_i}$ are the induced coordinates on $T^*Q$, called \textbf{extensive parameters}, 
then the (adjusted) Gibbs form is
$$
     \alpha = p_idx^i
$$
     and
\begin{equation}    \label{eq:td2.6}
  \omega = d\alpha = dp_i \wedge dx^i     
	\qquad (x^i =: x'^i \scirc \pi)
\end{equation}

\item[]
The special phase space (5) possesses a natural function:

\item[$\bullet$]
The \textbf{universal function of internal energy} is the function $U \in\FF M$, i.e., a map $T^*Q\to\mathbb R$, that can be expressed in the above linear coordinates as
$$
            U = p_ix^i
$$
Internal energy is a coordinate-independent concept and can be defined as 
\begin{equation}    \label{eq:td2.7}
  U(p) =: p(V_{\pi(p)})
\end{equation}
where $V \in\XX Q$ is the natural Liouville vector field on the
configuration space, and $p$, as a point of $T^*Q$,
is a differential form on $Q$.

\item[]
Since $dU=p_i\,dx^i+x^i\,dp_i$, the Gibbs form may be written as
\begin{equation}    \label{eq:td2.8}
  \alpha = -x^i\,dp_i + dU
\end{equation}
so that the Gibbs relation (\ref{eq:td2.2}) assumes the well-known textbook form:
$$
                \imath^* \, (x^i\,dp_i - dU) = 0
$$
i.e., $(x^i \scirc \imath) \; d(p_i \scirc \imath ) -d(U\scirc\imath) =0 $ \\


%


\vbox{\large 
\begin{center}
\hrule
\textbf{\large Thermodynamical Mandala}
\end{center}
\large
$$
\large
\begin{tikzpicture}[baseline=-0.4ex]
\tikzset{myptr/.style={decoration={markings,mark=at position 1 with %
    {\arrow[scale=3,>=stealth]{>}}},postaction={decorate}}}
    \matrix (m) [ matrix of math nodes,
                         row sep=2.5em,
                         column sep=3em,
                         text height=3ex, text depth=2ex] 
 {
             \quad  M \quad    
          & \quad U   \quad   
          & \quad \alpha  \quad   
          & \quad \omega  \quad   
          & \quad 0   \quad  \\
\quad  \Lambda \quad    
          & \quad   u   \quad   
          & \quad   0   \quad   
          & \quad   0   \quad  
          &   {~}\\
  };

    \path[-stealth]   (m-1-3) edge node[above] {$d$}   (m-1-4);
    \path[-stealth]   (m-1-4) edge node[above] {$d$}     (m-1-5);

    \draw[-stealth]        (m-2-1) edge node[right] {$\imath$}  (m-1-1);
    \path[stealth-]        (m-2-2) edge node[right] {$\imath^*$}  (m-1-2);
    \path[stealth-]        (m-2-3) edge node[right] {$\imath^*$}  (m-1-3);
    \path[stealth-]        (m-2-4) edge node[right] {$\imath^*$}  (m-1-4);

\node at (m-2-1.south) [below=0pt, color=black] {\smalll\sf {EQUATION} };
\node at (m-2-1.south) [below=11pt, color=black] {\smalll\sf {OF STATE} };

\node at (m-2-2.south) [below=0pt, color=black] {\smalll\sf {FUNDAMENTAL} };
\node at (m-2-2.south) [below=11pt, color=black] {\smalll\sf {RELATION} };

\node at (m-2-3.south) [below=0pt, color=black] {\smalll\sf { GIBBS-DUHEM} };
\node at (m-2-3.south) [below=11pt, color=black] {\smalll\sf {RELATIONS} };

\node at (m-2-4.south) [below=0pt, color=black] {\smalll\sf { MAXWELL~~ }};
\node at (m-2-4.south) [below=11pt, color=black] {\smalll\sf {definitions~~ }};

\node at (m-2-5.south) [below=7pt, left=2pt, color=black] {\smalll\sf { MAXWELL }};
\node at (m-2-5.south) [below=18pt, left=1pt, color=black] {\smalll\sf {IDENTITIES }};

\end{tikzpicture}   
$$
\vspace{-.5in}
\hrule
}

\vspace{.5in}

\item[$\bullet$]
\textbf{Fundamental relation} is an alternative way to select or describe a system $\Lambda\inn M$ and uses function $U$.
Let $\Lambda$ be a manifold with a scalar function set on it:
\[
            u: \Lambda \ \to \mathbb R
\]

\noindent
{\bf Problem:} 
Find an embedding $\imath:\Lambda\to M$ such that $u$ will match with $U$ along the embedding. i.e.,
\begin{equation}    \label{eq:td2.9}
  \imath^* U = u
\end{equation}
%
%

Equation \eqref{eq:td2.9} will be called the {\bf fundamental relation} and $u$ -- the {\bf defining function}.
The defining function $u$ should not be confused with the universal function of internal energy $U\in\FF M$.
The function $u\scirc \imath\equiv U|_\Lambda$ is the {\bf internal energy of the system} $\imath:\Lambda\to M$.

\item[]
It is easy to check that:

\begin{corollary}    \label{cor:td2.2}
The solutions to the equation of state \eqref{eq:td2.9} satisfy
automatically Gibbs-Duhem relation \eqref{eq:td2.3}.
\end{corollary}

\item[]
In practical applications, the above is realized as follows:
Let $\rho$ be a
projection from $M$ onto some
manifold $D$ of dim $D\leq {1\over 2} \dim M$.
We can set a problem: for a given function $u\in \FF D$, find the
submanifold $\imath:\; \Lambda \to M$ such that
\begin{equation}    \label{eq:td2.9'}
  \imath^* U = (\rho \scirc \imath )^* u
			\tag{9$'$}
\end{equation}
The basic example is that of $D=Q$ and $\rho=\pi$ (cf. \eqref{eq:td2.5}). 
\\

\begin{figure}[h]
\[
  \includegraphics[width=3.7in]{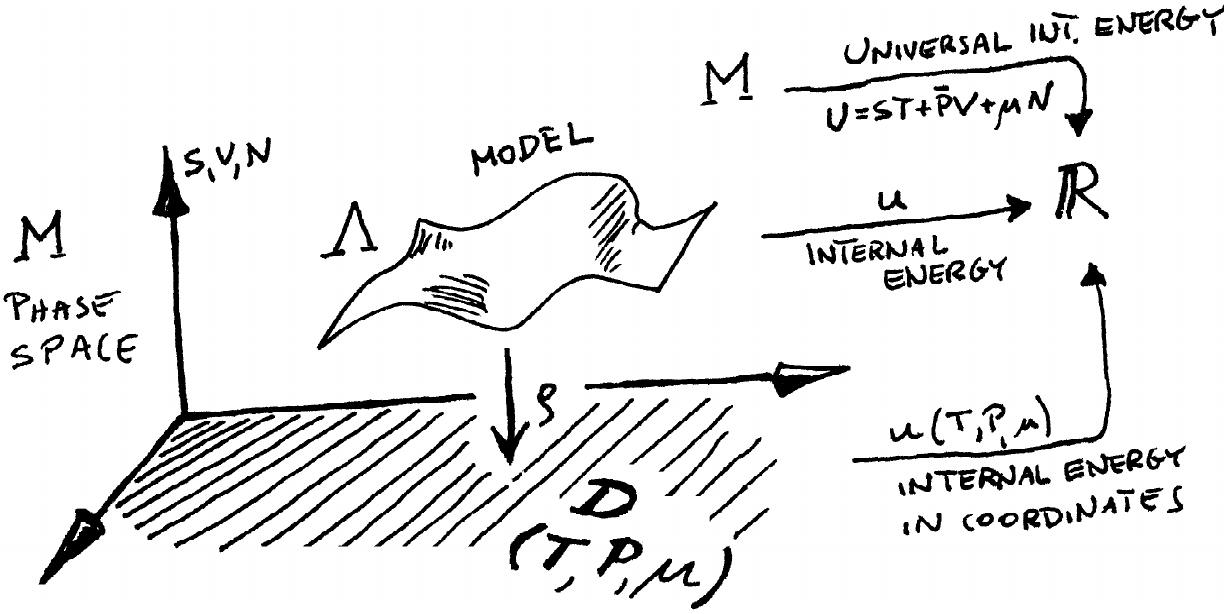}
\]
\caption{The fundamental relation}
\end{figure}

\hrule
\begin{equation}
\begin{tikzpicture}[baseline=-0.8ex]
    \matrix (m) [ matrix of math nodes,
                         row sep=3em,
                         column sep=4em,
                         text height=4ex, text depth=3ex] 
 {
             \quad  \Lambda \quad    
          & \quad M   \quad   
          & \quad \mathbb R  \quad    \\
\quad  ~ \quad    
          & \quad   D   \quad   
          & \quad   \mathbb R   \quad   \\
  };
    \path[-stealth]   (m-1-1) edge node[above] {$\imath$}  (m-1-2);
    \path[-stealth]   (m-1-2) edge node[above] {$U$}  (m-1-3);
    \path[-stealth]   (m-2-2) edge node[above] {$u$}  (m-2-3);

    \path[-stealth]        (m-1-2) edge node[right] {$\varrho$}  (m-2-2);
    \path[stealth-]        (m-1-3) edge node[right] {\rm id}  (m-2-3);
    \path[-stealth]        (m-1-1) edge [bend right] node[above] {$\qquad\varrho\circ\imath$}  (m-2-2);
\end{tikzpicture}   
\end{equation}
\hrule

\end{enumerate}


\leftskip=0in


\section{Standard model -- an example}
\label{sec:td.3}

The standard model of thermodynamics is based on a 6-dimensional space:
\begin{equation}    \label{eq:td.1}
\begin{alignedat}{2}
  \noalign{\hbox{\textsc{Intensive Variables}}}
  x^1 &= T   &&\hbox{ (temperature)} \\
  x^2 &= -P \equiv \mP &&\hbox{ (pressure)}  \\
  x^3 &= \mu &&\hbox{ (chemical potential)}
\end{alignedat}
\hbox to .7in{}
\begin{alignedat}{2}
  \noalign{\hbox{\textsc{Extensive Variables}}}
  p^1 &= S &&\hbox{(entropy)} \\
  p^2 &= V &&\hbox{(volume)}  \\
  p^3 &= N &&\hbox{(number of molecules)}
\end{alignedat}
\end{equation}
Extensive variables are additive and are proportional to the size of
the system.  Intensive variables are independent of the size of the
system and specify a local property.
The differential Gibbs form is thus
\begin{equation}    \label{eq:td3.2}
\begin{aligned}
  \alpha &= S\,dT + V\,d\mP + N\,d\mu  \\
         &= -T\,dS - \mP\,dV - \mu\,dN + dU  
\end{aligned} 
\end{equation}
where
\begin{equation}    \label{eq:td3.3}
  U\ = \ TS+\mP V+\mu N
\end{equation}
The symplectic form on $M$ is
\begin{equation}    \label{eq:td3.4}
  d \alpha = dS\wedge dT +dV\wedge d\mP + dN\wedge d\mu
\end{equation}
The embedding $\imath$ of the system \eqref{eq:td2.2} may be set by ``constraints'' of the type $PV=\hbox{const}\cdot T$ (``ideal gas equation;'' 
see Table (\ref{tbl:td.1}) for other theories), or, equivalently, by the fundamental relation \eqref{eq:td2.9}.  
For example, let $Q$ be the subspace of $M$ spanned by directions $(T,P,\mu)$, and let the defining function $u\in\FF Q$ be chosen, i.e. $u=u(T,P,\mu)$.
Denote
\[
\begin{aligned}
  S\scirc \imath   &= S(T,P,\mu) \\
  V\scirc \imath   &= V(T,P,\mu) \\
  N\scirc\ \imath  &= N(T,P,\mu)  
\end{aligned}
\]
Then the fundamental relation (2.10) is an equation for an embedding
$\imath$ such that the following is satisfied
\begin{equation}    \label{eq:td3.5}
  u(T,P,\mu) = S(T,P,\mu)\cdot T + V(T,P,\mu)
	\cdot P + N(T,P,\mu)\cdot\mu
\end{equation}
which agrees with the standard textbook definitions.

\begin{table}
\hrule
\[
\begin{alignedat}{2}
  &\hbox{ideal gas:} 
	&\hbox to .2in{}&pV = RTN \\
  &\hbox{van der Waals:} 
	&&\left( p + \frac{a}{V^2} \right)\;( V-b) = RTN \\
  &\hbox{Berthelot:}
	&&\left( p + \frac{a}{TV^2} \right)\;( V-b) = RTN \\
  &\hbox{Dieterici:}
	&&p(V-b)\; e^{\frac{a}{RTV}}  = RTN \\
  &\hbox{Onnes:}
	&&pV = NRT\; \left(1+ B(T)\frac{a}{V} 
	+ C(T)\left(\frac{a}{V}\right)^2 + \dots \right)
	\quad \left( \begin{gathered} \hbox{\small Virial} \\[-2pt]
		\hbox{\small Expansion} \end{gathered} \right)
\end{alignedat}
\]
\caption{Some theories of PTE (see \cite{KP} p.\ 158)}
\label{tbl:td.1}
\vskip.1in
\hrule
\end{table}

\section{Seven (or $\infty$) potentials of PTE}
\label{sec:td.4}

The Gibbs form $\alpha$ of the phase space can always be expressed in the canonical coordinates as $\alpha=p_i\,dx^i$ (Darboux Theorem).
If for any reason we would like to have a term $x^{\widehat k} dp_{\widehat k}$ instead of $p_{\widehat k} dx^{\widehat k}$ for any $k$
(the hat above an index $k$ denotes that $k$ is fixed and no summation is intended), then the exterior derivative of $p_{\widehat k} x^{\widehat k}$ must appear in the expression of Gibbs form
\begin{equation}    \label{eq:td4.1}
  \alpha = p_i\,dx^i
          = \sum_{i\not= k} p_i\,dx^i  - x^{\widehat k}\,dp_{\widehat k}
            + d(p_{\widehat k}x^{\widehat k})
\end{equation}
Such a coordinate-flip can be done for any pair of conjugated
canonical variables on $2n$-dimensional $M$.
Consequently, every subset $J$ of $I=\{1,2,\ldots,n\}$ determines
a function:
\begin{equation}    \label{eq:td4.2}
  f_J = \sum_{k\in J} \; p_k \, x^k
\end{equation}
Clearly, there are $2^n$ independent functions of this type. In
particular, we assume $f_\Phi=0$.  Function $f_J$ will be called
the $J$-{\it th thermodynamic potential}.  The Gibbs form can be
expressed in a form involving the $J$-th potential
\begin{equation}    \label{eq:td4.3}
  \alpha = \sum_{i\in I-J} p_i\,dx^i
            - \sum_{k\in J} x^k\,dp_k + df_J
\end{equation}
For the 6 coordinates of the standard model \eqref{eq:td.1}, 
we have 7 non-zero potentials that form a lattice structure isomorphic with the lattice of subsets of the set $I=\{1,2,3\}$
\def\rrr{\phantom{{}^{L}}}
\begin{center}
\begin{tikzpicture}
    \tikzstyle{all nodes}=[inner sep=4pt]
    \draw node(t) at (0,0)   {\fbox{$\rrr U=TS+\bar P V+\mu N\rrr$}}
          node(a) at (-3.7,-1.5)  {\fbox{$\rrr F=\mP V+\mu N\rrr $}}
          node(b) at (0,-1.5)   {\fbox{$\rrr I=\mP V+TS\rrr $}}
          node(c) at (3.7,-1.5)   {\fbox{$\rrr H = TS+\mu N\rrr $}}
          node(A) at (-3.7,-3.5) {\fbox{$\rrr \Omega=\mP V\rrr $}}
          node(B) at (0,-3.5)  {\fbox{$\rrr G=\mu N\rrr $}}
          node(C) at (3.7,-3.5) {\fbox{$\rrr \Gamma=TS\rrr $}}
          node(d) at (0,-5.2) {\fbox{$\rrr 0\rrr$}};
     \draw (t)--(a) (t)--(b) (t)--(c)
           (a)--(A) (a)--(B) (b)--(A) (b)--(C) (c)--(B) (c)--(C)
           (A)--(d) (B)--(d) (C)--(d);
\end{tikzpicture}
\end{center}

\noindent
where some of the functions are traditionally called 

\vskip.1in
\begin{tabular}{cclcccl}
$U$ &- & internal energy          &&$\Gamma$  &- & (no name) \\
$F$ &- & Helmholtz free energy    &&$G$    &- & Gibbs potential \\
$I$ &- & (no name)                &&$\Omega$  &- & grand potential \\
$H$ &- & enthalpy                 && 0     &- & just zero!
                             (trivial potential) 
\end{tabular}

\vskip.1in
Different potentials appear naturally in different situations.
Typically, the choice of coordinates determines convenient potential to be used. 
For example, if the coordinates on $\Lambda\inn M$ are induced from $(T,V,N)$ on $M$, 
then potential $F$ finds its natural application since
the Gibbs form \eqref{eq:gibbs} may be rewritten as 
\begin{equation}    \label{eq:td4.4}
  \alpha = S\,dT -\mP\,dV - \mu\,dN + dF 
\end{equation}
Thus
\begin{equation}    \label{eq:td4.5}
  \imath^*\alpha = S(T,V,N)\,dT - \mP(T,V,N)\,dV 
	- \mu(T,V,N)\,dN + dF(T,V,N)
\end{equation}
so that Gibbs-Duhem relation $\imath^*\alpha=0$ reduces to
\begin{equation}    \label{eq:td4.6}
\begin{aligned}
\imath^*S          =   S(T,V,N)    &\ = \ - {{\q F\scirc \imath} \over \q T} \midx_{V,N} \\[3pt]
\imath^*\mP          =   \mP(T,V,N)     &\ = \ \phantom{-} {{\q F\scirc \imath} \over \q V} \midx_{T,N} \\[3pt]
\imath^*\mu      =  \mu(T,V,N) &\ = \  \phantom{-}  {{\q F\scirc \imath} \over \q N} \midx_{T,V} 
\end{aligned}
\end{equation}
Note that \eqref{eq:td4.6} these can be viewed as defining relations for the PTE variables.
To see how the choice of coordinates suggests a convenient potential, consider, for instance, how the entropy function $S$ can be related to other three potentials
\begin{equation}    \label{eq:td4.7}
  \imath^*S \ = \ -{{\q G\scirc \imath} \over \q T} \big|_{P,N}
    \  = \  -{{\q \Omega\scirc \imath} \over \q T} \big|_{N,\mu}
    \  = \  -{{\q F\scirc \imath} \over \q T} \big|_{\mu,P} 
\end{equation}
By analogy, all of the $3\cdot 7$  such relations can be expressed.
\\

\noindent
{\bf Potentials without coordinates.}
The linearity of the special phase space allows one to define
thermodynamical potentials in a coordinate-free way.  Indeed, each
decomposition $\mathcal A$ of the space $Q$ into two subspaces
$$
        Q=A\oplus B \qquad \pi_A: Q\to A
$$
determines a potential function $f_\mathcal A$ on $T^*Q$ which at point
$p\in T^*Q$ assumes value
\begin{equation}    \label{eq:td4.8}
  f_\mathcal A(p) = p((\pi_A)_* V_{\pi(p)})
\end{equation}
where $V_{\pi(p)}$ is the Liouville vector evaluated a point $\pi(p)$.
Function $f_\mathcal A$ may be called the thermodynamical potential for the splitting $\mathcal A$.

This defines an infinite lattice $\LL$ of thermodynamical potentials
with the structure induced from the natural lattice of subspaces $A$ and $B$ in the linear space $Q$.  
The maximal element of the lattice, $U$, corresponds to the trivial decomposition $Q=Q\oplus\{\bf 0\}$.  
The minimal element, the constant null function, corresponds to
$Q=\{ {\mathbf 0}\}\oplus Q$.  
The Boolean lattice in Figure 2 is a finite sublattice of $\LL$.

\section{Gauge interpretation of PTE}
\label{sec:td.5}

Let $\{E,\pi, M\}$ be a fiber bundle over the thermodynamical phase
space $M$ with real line $\mathbb R$ as the typical fiber, and with a
projection
$$
   \pi: E \ \to \ M
$$
Define {\it Gibbs connection} $\nabla$ as a vector-valued exterior
form
\begin{equation}    \label{eq:td5.1}
  \nabla = \q_u\otimes\alpha
\end{equation}
The curvature of the connection is
\begin{equation}    \label{eq:td5.2}
  \hbox{Curv}\;(\nabla) = \q_u \otimes d\alpha = \q_u\otimes\omega
\end{equation}
A system can be redefined as a submanifold $\Lambda\inn M$ such that
the curvature form on the subbundle $\{\pi^{-1}(\Lambda), \pi,\Lambda\}$ vanishes.

\[
  \includegraphics[width=2in]{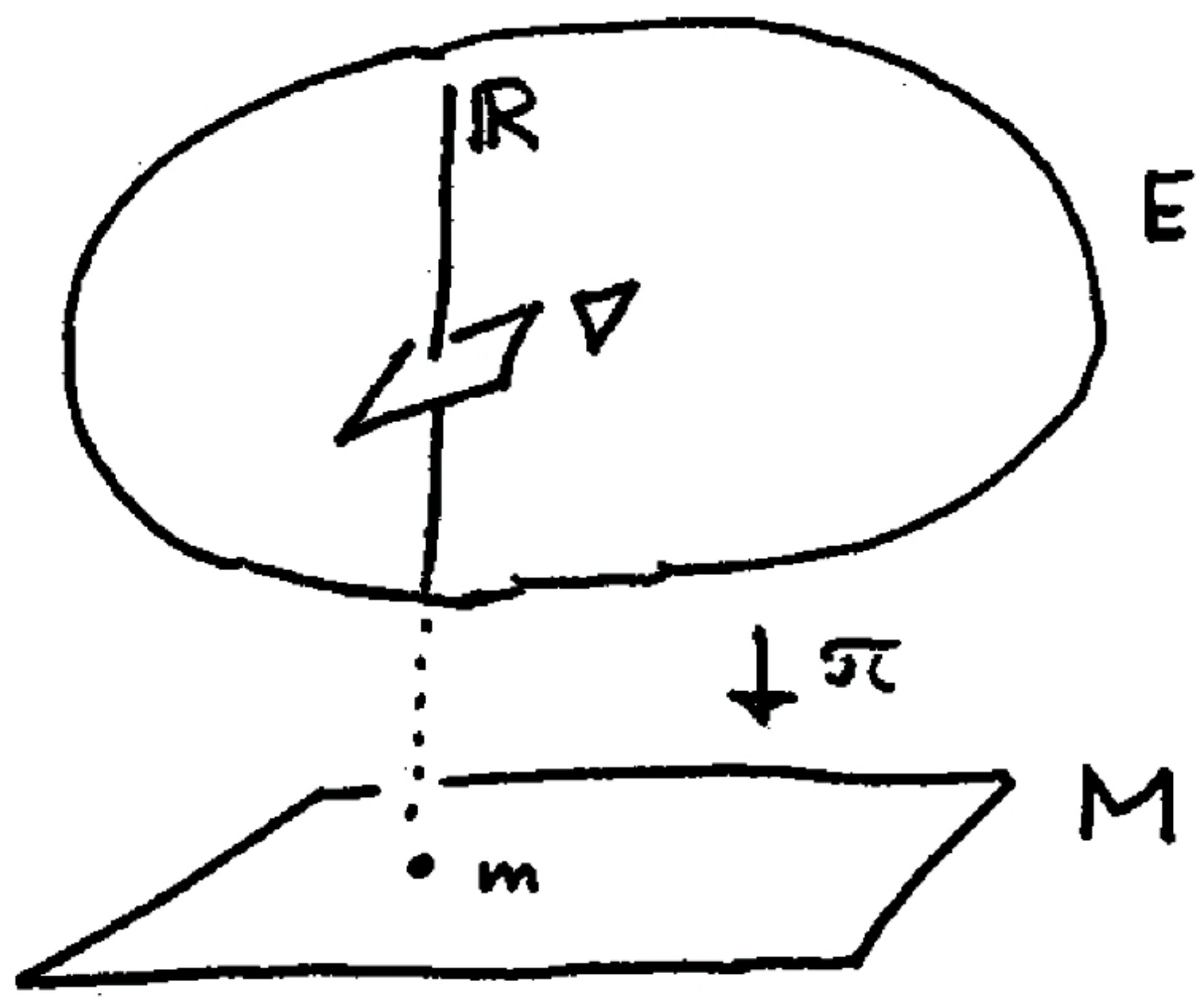}
\]

A potential can be viewed now as a section of the fiber bundle
\begin{equation}    \label{eq:td5.3}
  f\in \hbox{Sec}\,(E,\pi,M) \qquad f: M\to E, 
	\qquad \pi\scirc f = \hbox{\rm id}_M
\end{equation}
Each such section induces a differential one-form $\alpha_f=\nabla f$ on
$M$ (see Equation 20) defined by its value on any vertical vector
$v\in TM$
$$
\alpha_f(v)=\nabla_v f
$$
Thus the known potentials correspond to convenient ``choices of gauge'' for the Gibbs connection $\alpha$.
In this context the holonomy group has the meaning of work required to perform a cyclic process.
Indeed, if $\gamma:I\to M$ is a loop process in the phase space, then the curve $\tilde\gamma$ lifted to space $E$ by the connection ends in the same fiber $\pi^{-1}\gamma(0)$, but possibly in a point different from the initial, and then $\tilde\gamma(1)-\tilde\gamma(0)=\Delta U$.

\section{Comparison with classical mechanics}
\label{sg.6}

We conclude with a few remarks on classical mechanics versus
phenomenological thermodynamics. It became customary to stress the
similarity between structure of PTE and that of classical mechanics
(CM), whereas they should be rather contrasted.
\\

\noindent 
{\bf Canonical Structure.}
Although both PTE and CM rely essentially on a manifold with the closed differential biform $\omega=d\alpha$ of the (pre-)symplectic structure (Poisson bracket), yet the one-form $\alpha$ plays a different role in each formalism:

\begin{enumerate}
\item[$\bullet$]
in CM, form $\alpha$ is defined up to an external derivative on any
function on the manifold of the phase space, i.e. the gauge
$\alpha \to \alpha' = \alpha + df$ does not change the classical equation
of motion $X\jjj d\alpha = dH$, since $d\alpha = d\alpha'$

\item[$\bullet$]
in PTE things seem to be quite opposite: one of the
one-forms (the Duhem-Gibbs form) generating the symplectic
structure, namely Gibbs form $\alpha$, is {\it distinguished},
and this fact is crucial for the theory.
The equation for the submanifold $\Lambda$ of admissible states, $\imath^*\alpha= 0$,
is not invariant under the gauge $\alpha\to\alpha+df$.
(Only the implied Maxwell identity $\imath^* d\alpha =0$ {\it is}
invariant).  This fact is the source of importance of thermodynamical
potentials in PTE.

\end{enumerate}
%

\noindent
{\bf Legendre-- versus Gauge Transformations.}
The appearance of different potentials in PTE is often compared to
Legendre transformation of CM.

\begin{enumerate}
\item[$\bullet$]
In CM, the Legendre transformation is essentially the map between tangent and cotangent bundles over configuration space, $TQ\to T^*Q$,  so that the induced map transforms the ``messy'' symplectic form $d(\q_vL)\wedge dx$ defined by the Lagrangian $L$ on $TQ$ into the canonical form $dp\wedge dx$ on $T^*M$ (Darboux variables).

\item[$\bullet$]
This context does not appear in PTE, where the pre-symplectic
form $d\alpha = dS\wedge dT +dV\wedge d\mP+ dN\wedge d\mu$ has already the canonical form. 
The potentials appear when one would like to express the {\it same} form $\alpha$ in another basis of differentials on $M$
(like e.g. $T\,dS= -S\,dT+d\,(TS)$).

\end{enumerate}
%


\begin{remark}
If coordinates $S,V,N$ (i.e., extensive variables) are chosen to be used on $\Lambda$, then the ``method'' gives
\[
  \alpha \quad = \quad - \ \underset{\hbox{\tiny change}}{\underset{\hbox{\tiny heat}}{\underbrace{T\,dS}_{~\dbar Q\phantom{\big|} }}}
	 \ \ - \ \   {\underset{\hbox{\tiny work}}{\underbrace{\mP\,dV - \mu\,dN}_{~\dbar W\phantom{\big|}}}
	\ \ + \ \ \underset{\hbox{\tiny energy}}{\underset{\hbox{\tiny internal}}{\underbrace{d(TS+\mP V+\mu N)}_{d U\phantom{\big|}}}}}
\]
\end{remark}

\noindent
$\alpha \big|_\Lambda = 0$ is equivalent to energy conservation (for any ``theory'' $\Lambda$), namely
\[
  \underset{\hbox{\tiny to the system}}{\underset{\hbox{\tiny transferred}}{\underset{\hbox{\tiny heat}}{dQ\phantom{\big|}}}}
	\ = \  \underset{\hbox{\tiny the system}}{\underset{\hbox{\tiny done by}}{\underset{\hbox{\tiny work}}{dW\phantom{\big|}}}}
	+ \underset{\hbox{\tiny energy}}{\underset{\hbox{\tiny internal}}{\underset{\hbox{\tiny change of}}{dU\phantom{\big|}}}}
\hbox to .6in{}
\]

\begin{remark}
Some authors, in an attempt to geometrize thermodynamics, consider
\[
  \theta = T\,dS - P\, dV + \mu\, dN - dU
\]
as the \textbf{basic}  form on an \textbf{odd-dimensional} manifold with coordinates $(T,S,p,V,\mu,N,u)$, $u$ as an independent variable!
\end{remark}

%


\section*{Appendix A:  \ Magic thermodynamical cube}

\begin{figure}[h!]
\centering
\includegraphics[width=2.2in]{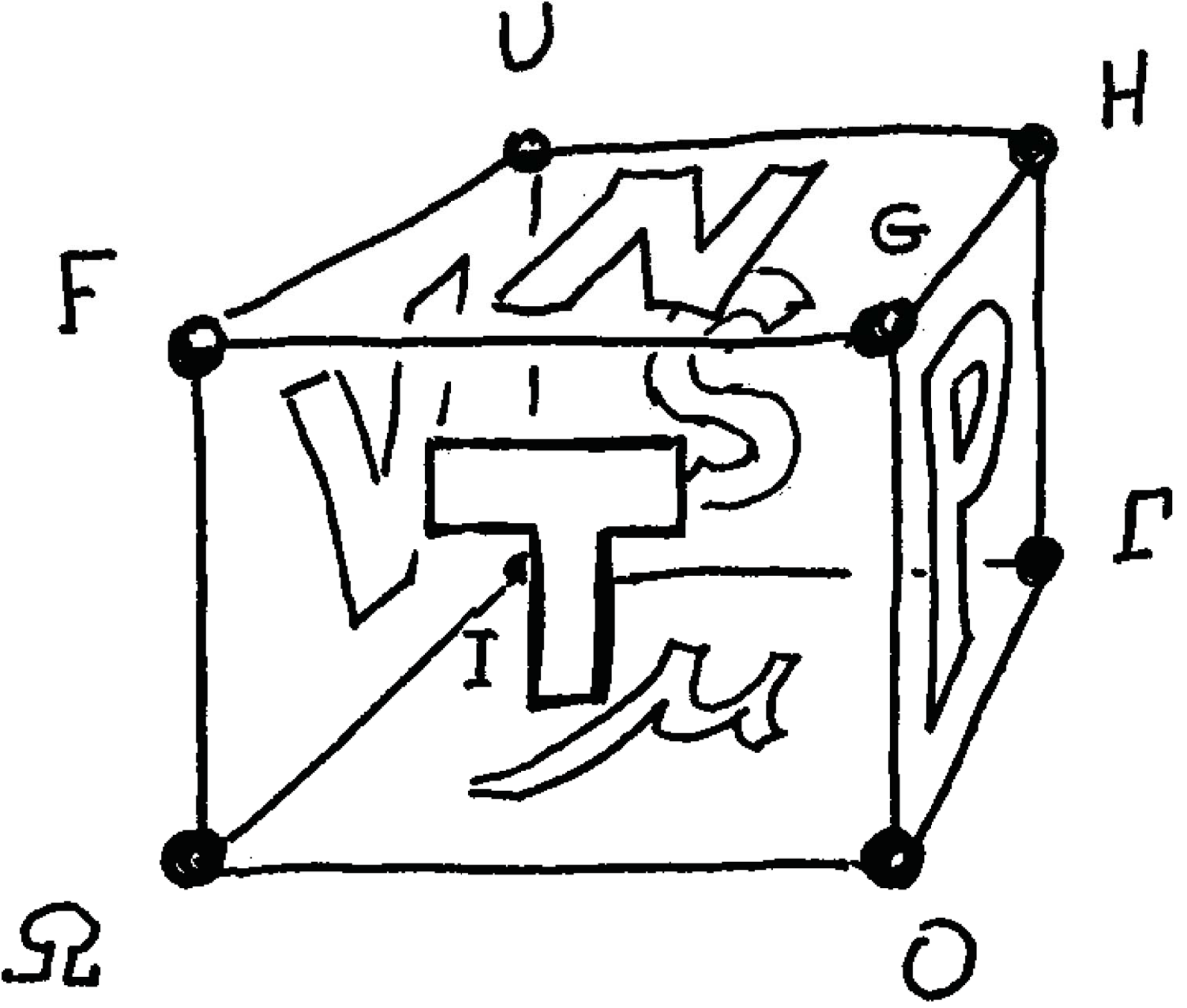}\qquad\quad
\includegraphics[width=1.5in]{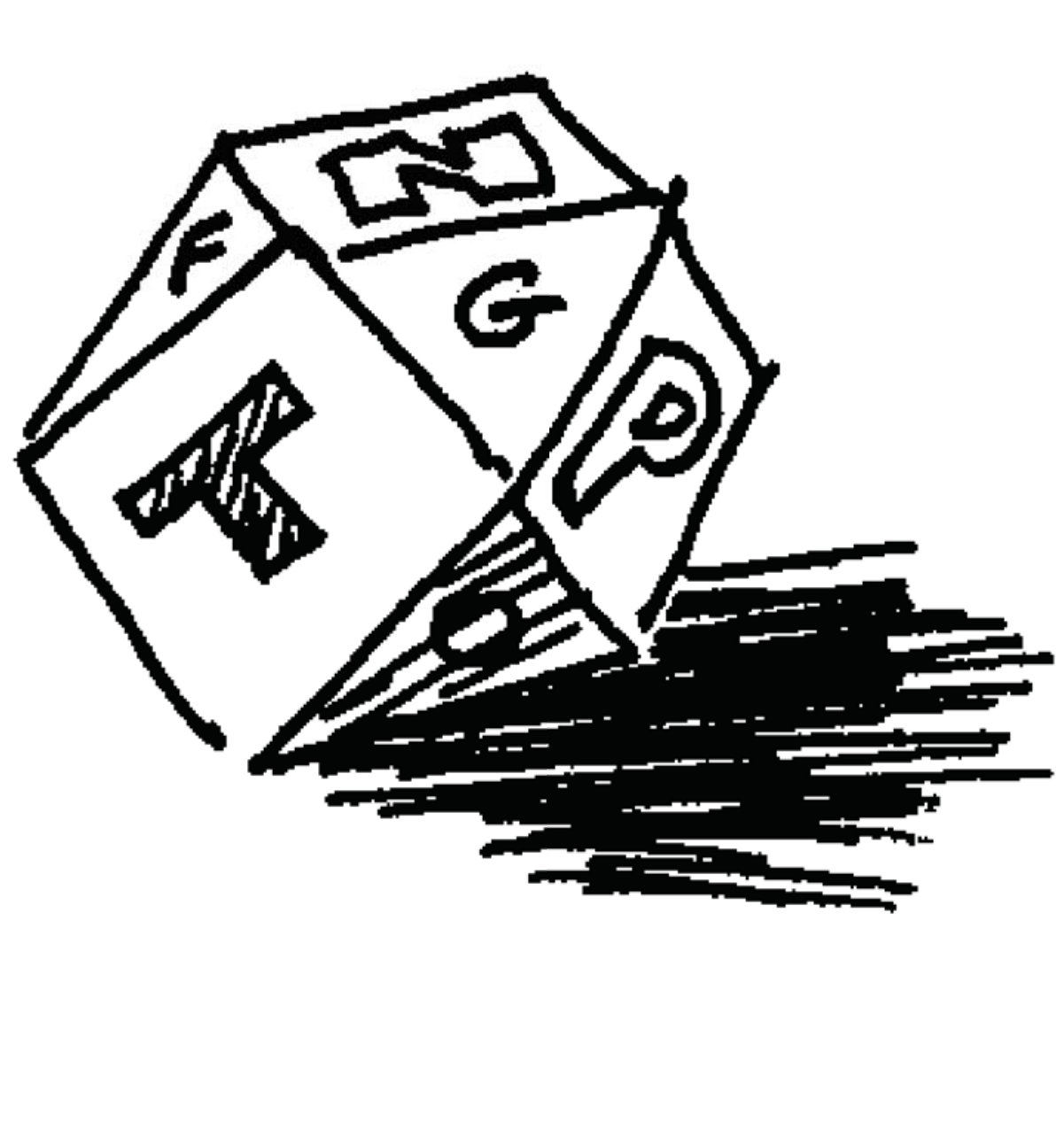}
\caption{Thermodynamic cube}
\label{fig:th-cube}
\end{figure}

\def\cubeone{\square}
\def\cubetwo{\square}
\def\cubethree{\square}
\def\cubefour{\square}
\def\cubefive{\square}
\def\cubesix{\square}

{\bf How to Read It}

\def\cubeA{%
\begin{tikzpicture}[baseline=.59ex, scale=.45]
    \draw   (0.0,0.0) -- (0.5, 0.5) -- (0.5, 1.4) -- (0.0, 1.0) -- cycle;  
    \draw (1.0,0.0) -- (1.4, 0.5)-- (1.4, 1.4)--(1.0, 1.0)--cycle;  
    \draw  (0.0,0.0)  -- (1.0, 0.0)-- (1.4, 0.5)--(0.5, 0.5)--cycle;
    \draw  (0.0,1.0) -- (1.0, 1.0)-- (1.4, 1.4)--(0.5, 1.4)--cycle;
    \draw  (.5,1.4) -- (1.0, 0);
       \draw[fill=black]  (0.5,1.4) circle (3pt);
       \draw[fill=black]  (1,0) circle (3.75pt);
    \end{tikzpicture}
}

\def\cubeB{%
\begin{tikzpicture}[baseline=.59ex, scale=.45]
    \draw   (0.0,0.0) -- (0.5, 0.5) -- (0.5, 1.4) -- (0.0, 1.0) -- cycle;  
    \draw (1.0,0.0) -- (1.4, 0.5)-- (1.4, 1.4)--(1.0, 1.0)--cycle;  
    \draw  (0.0,0.0)  -- (1.0, 0.0)-- (1.4, 0.5)--(0.5, 0.5)--cycle;
    \draw  (0.0,1.0) -- (1.0, 1.0)-- (1.4, 1.4)--(0.5, 1.4)--cycle;
    \draw  (0,1) -- (1.0, 0);
       \draw[fill=black]  (0,1) circle (3.75pt);
       \draw[fill=black]  (1,0) circle (3.75pt);
    \end{tikzpicture}
}

\def\cubeBB{%
\begin{tikzpicture}[baseline=.59ex, scale=.45]
    \draw   (0.0,0.0) -- (0.5, 0.5) -- (0.5, 1.4) -- (0.0, 1.0) -- cycle;  
    \draw (1.0,0.0) -- (1.4, 0.5)-- (1.4, 1.4)--(1.0, 1.0)--cycle;  
    \draw  (0.0,0.0)  -- (1.0, 0.0)-- (1.4, 0.5)--(0.5, 0.5)--cycle;
    \draw  (0.0,1.0) -- (1.0, 1.0)-- (1.4, 1.4)--(0.5, 1.4)--cycle;
    \draw  (0,0) -- (1.0, 1.0);
       \draw[fill=black]  (0,0) circle (3.75pt);
       \draw[fill=black]  (1,1) circle (3.75pt);
    \end{tikzpicture}
}

\def\cubeCC{\begin{tikzpicture}[baseline=.59ex, scale=.45]
    \draw   (0.0,0.0) -- (0.5, 0.5) -- (0.5, 1.4) -- (0.0, 1.0) -- cycle;  
    \draw (1.0,0.0) -- (1.4, 0.5)-- (1.4, 1.4)--(1.0, 1.0)--cycle;  
    \draw  (0.0,0.0)  -- (1.0, 0.0)-- (1.4, 0.5)--(0.5, 0.5)--cycle;
    \draw  (0.0,1.0) -- (1.0, 1.0)-- (1.4, 1.4)--(0.5, 1.4)--cycle;
    \draw  (0,0) -- (0, 1.0);
       \draw[fill=black]  (0,0) circle (3.75pt);
       \draw[fill=black]  (0,1) circle (3.75pt);
    \end{tikzpicture}
}

\def\cubeCC{\begin{tikzpicture}[baseline=.59ex, scale=.45]
    \draw   (0.0,0.0) -- (0.5, 0.5) -- (0.5, 1.4) -- (0.0, 1.0) -- cycle;  
    \draw (1.0,0.0) -- (1.4, 0.5)-- (1.4, 1.4)--(1.0, 1.0)--cycle;  
    \draw  (0.0,0.0)  -- (1.0, 0.0)-- (1.4, 0.5)--(0.5, 0.5)--cycle;
    \draw  (0.0,1.0) -- (1.0, 1.0)-- (1.4, 1.4)--(0.5, 1.4)--cycle;
    \draw  (0,0) -- (0, 1.0);
       \draw[fill=black]  (0,0) circle (3.75pt);
       \draw[fill=black]  (0,1) circle (3.75pt);
    \end{tikzpicture}
}

\def\cubeDD{\begin{tikzpicture}[baseline=.59ex, scale=.45]
    \draw   (0.0,0.0) -- (0.5, 0.5) -- (0.5, 1.4) -- (0.0, 1.0) -- cycle;  
    \draw (1.0,0.0) -- (1.4, 0.5)-- (1.4, 1.4)--(1.0, 1.0)--cycle;  
    \draw  (0.0,0.0)  -- (1.0, 0.0)-- (1.4, 0.5)--(0.5, 0.5)--cycle;
    \draw  (0.0,1.0) -- (1.0, 1.0)-- (1.4, 1.4)--(0.5, 1.4)--cycle;
    \draw  (0,0) -- (0, 1.0);
           \draw[fill=black]  (0,1) circle (3.75pt);
    \end{tikzpicture}
}

\def\cubeFF{\begin{tikzpicture}[baseline=.59ex, scale=.45]
    \draw  (0.0,0.0) -- (0.5, 0.5) -- (0.5, 1.4) -- (0.0, 1.0) -- cycle;  
    \draw (1.0,0.0) -- (1.4, 0.5)-- (1.4, 1.4)--(1.0, 1.0)--cycle;  
    \draw  (0.0,0.0)  -- (1.0, 0.0)-- (1.4, 0.5)--(0.5, 0.5)--cycle;
    \draw  (0.0,1.0) -- (1.0, 1.0)-- (1.4, 1.4)--(0.5, 1.4)--cycle;
    \draw[very thick, color=gray]  (0.1,0.6)  -- (1.0, 0.6)-- (1.35, 0.9)--(0.5, 0.9)--cycle;
    \end{tikzpicture}
}

\begin{enumerate}
\item[1.] $\displaystyle\sum\limits_{\rm Any \ cube\atop diagonal} \cubeA \ \ = \ \ U$ \qquad (sum of potential on any cube diagonal equals to $U$)\\

\item[2.]  \  \ $\displaystyle\sum \ \cubeB \ \ = \ \ \displaystyle\sum \ \cubeBB$ \qquad (Good for any face diagonals)
\item[3.] Two potentials on the same edge $\cubeCC$ differ by a product of the adjacent faces, for instance:
\marginpar{\centering       
            \includegraphics[width=.75in]{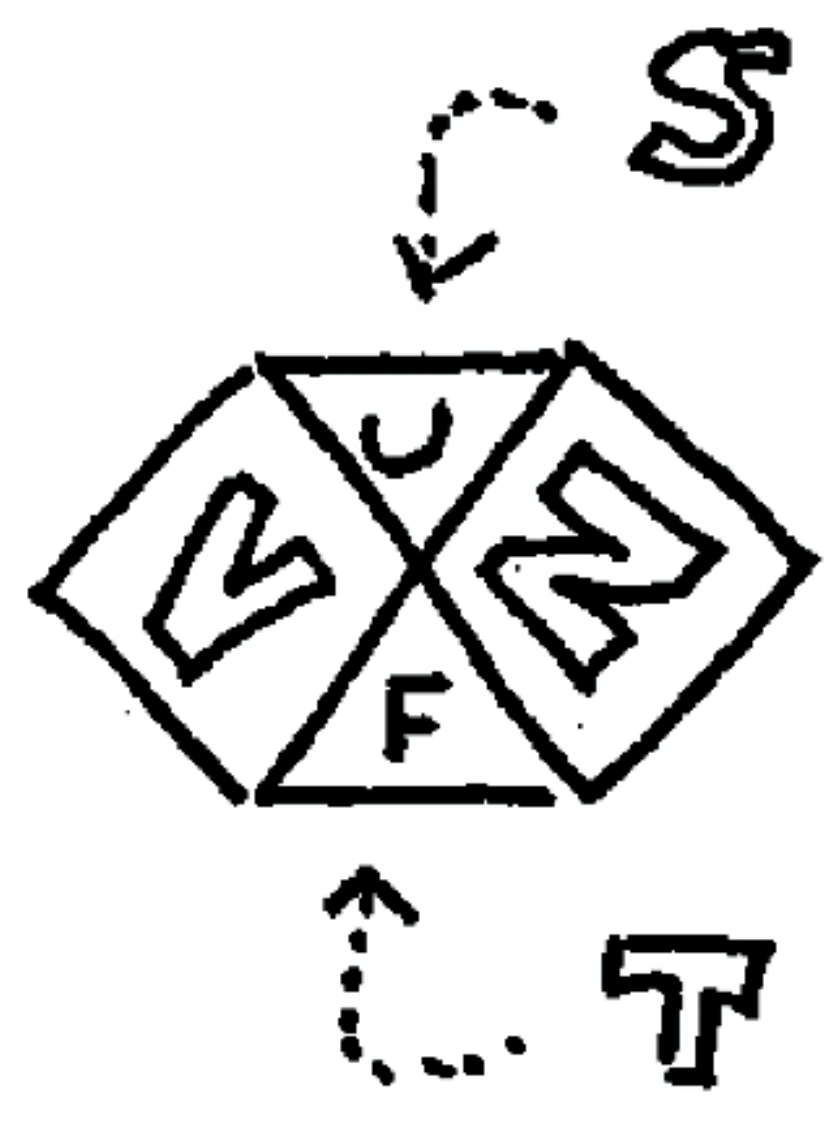}%
}
\[
  U-F = ST, \qquad F-\Omega = VT ,   \qquad F-G  = \mP V,  \ \ etc.
\]

\item[4.] 
{\bf Finding associated variables:}
Any three adjacent variables are (admissible) convenient to
describe a system. The potential in the corner and the three variables
are called {\it associated}.  
Choose a potential, say $F$.  Turn the cube so that the corresponding triangle
$F$ is facing you.  
\marginpar{\centering       
            \includegraphics[width=.75in]{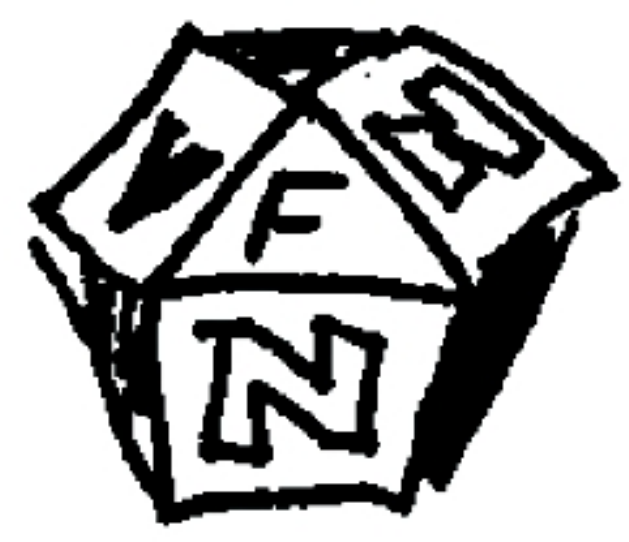}%
}
The associated variables
are those in the squares surrounding the triangle,  $V$, $N$, and $T$.
\[
  \cubeDD  \qquad {\rm Potential} \ F \  \hbox{and variables} \ \{T,N, V\}  
\]

\item[5.] 
{\bf Gibbs relations:} Each of the thermodynamical variables can be  defined in terms of
a partial derivative of a potential.  To find expression for a particular
variable (say $P$), turn the cube to see the opposite face ($V$).
The variable in question, $\mP \equiv -P$, is equal to the partial derivative 
of any visible potential (triangle) with respect to the variable 
in the square.
$$
-P \  = \ {{\q U} \over \q V} \bigg|_{S,N}
  \  = \  {{\q F} \over \q V} \bigg|_{T,N}
  \  = \  {{\q \Omega} \over \q V} \bigg|_{T,\mu}
  \  = \  {{\q I} \over \q V} \bigg|_{S,\mu}
$$
%
\marginpar{\centering       
            \includegraphics[width=.85in]{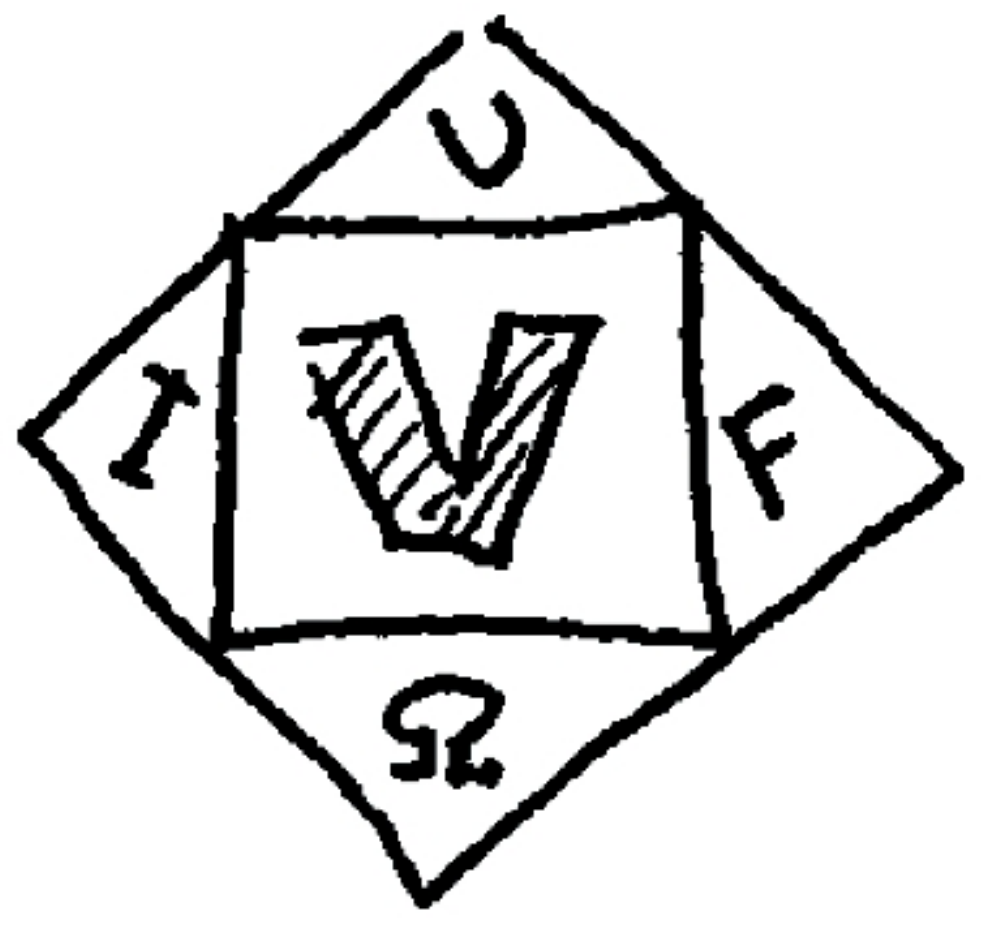}%
}
Notice that each derivative is taken in another system of variables, the
ones associated to the corresponding potential, and therefore different
variables are set constant.

%
%
%
\item[6.] {\bf Maxwell's identities:}  Any belt of four faces, e.g horizontal $\{ S,V,T,\mP \}$, gives Maxwell identities:
\[
 \cubeFF \qquad
 \dfrac{\q S}{\q V} \bigg|_{T,N}= \dfrac{\q P}{\q T}\bigg|_{V,N}\qquad \hbox{or}\qquad 
 \dfrac{\q S}{\q V} \bigg|_{T,\mu}= \dfrac{\q P}{\q T}\bigg|_{V,\mu}
\]
\end{enumerate}
%
\marginpar{\centering       
            \includegraphics[width=1.2in]{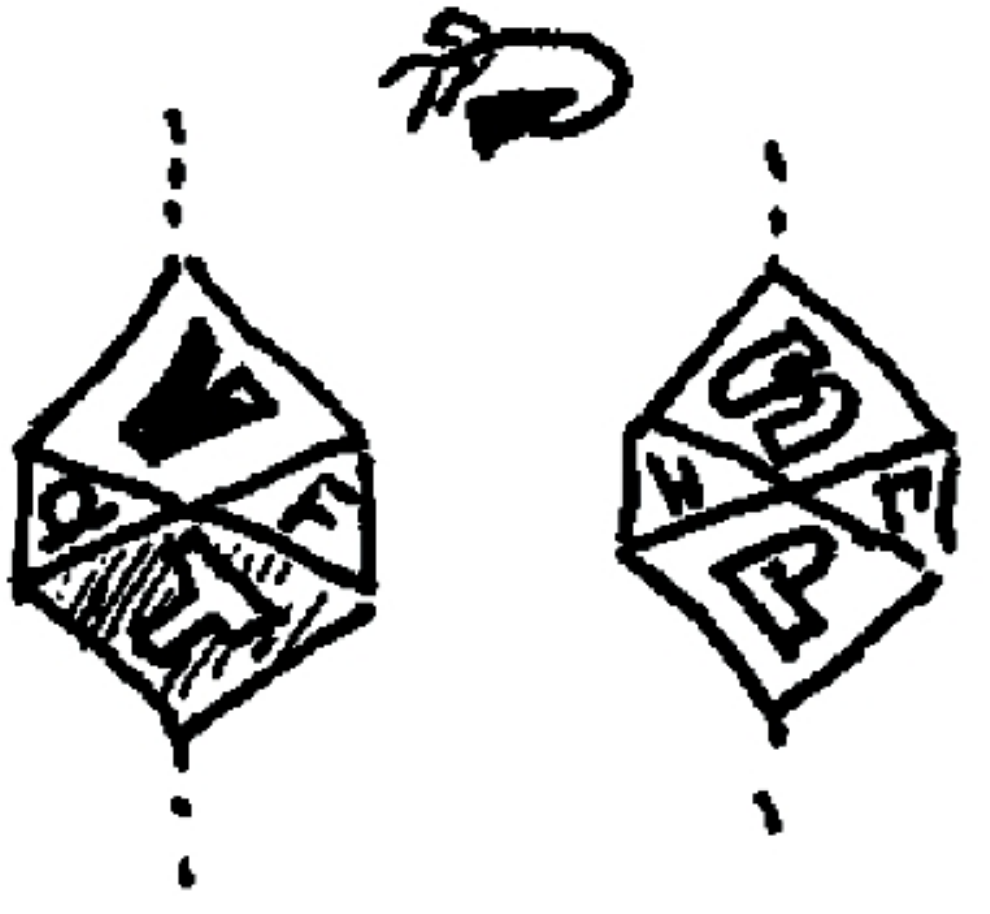}%
}

~

\section*{Appendix B:  Various identities in coordinates}

For the reader who wants to verify readings from the ``magic cube of thermodynamics'',
the basic identities are listed below.

\subsection*{TABLE 1. Potentials:}

$$
\begin{aligned}
       U &= TS +\mP V+\mu N  \\
       F &= \mP V+\mu N     \\
       I &= TS+\mP V        \\
       H &= TS+\mu N     
\end{aligned}
\qquad
\begin{aligned}
   \Gamma &=  TS       \\
        G &=  \mu N    \\
   \Omega &= \mP V       \\
        0 &=  0        
\end{aligned}
$$

\subsection*{TABLE 2.  Potential differences:}

$$
\begin{aligned}
U - F        &=  ST     \\
U - H        &=  V\mP    \\
U - I        &=  N\mu   \\
H - G        &=  ST     \\
H - \Gamma   &=  N\mu   \\
F - G        &=  V\mP    
\end{aligned}
\qquad
\begin{aligned}
F - \Omega   &=  \mu N  \\
I - \Gamma   &=  V\mP    \\
I - \Omega   &=  ST     \\
\Omega-0     &=  \mP V    \\
G - 0        &=  N\mu   \\
\Gamma-0     &=  ST     
\end{aligned}
$$

\subsection*{TABLE 3. Associated variables:}

$$
\begin{aligned}
U      &\quad\zzz\quad  ( S, V, N)     \\
F      &\quad\zzz\quad  ( T, V, N)     \\
I      &\quad\zzz\quad  ( S, V, \mu)   \\
H      &\quad\zzz\quad  ( S, P, N )    
\end{aligned}
\qquad
\begin{aligned}
\Gamma &\quad\zzz\quad  ( S, P, \mu)    \\
G      &\quad\zzz\quad  ( T, P, N  )   \\
\Omega &\quad\zzz\quad  (T, V, \mu)    \\
0      &\quad\zzz\quad  (T, P, \mu)    
\end{aligned}
$$

\begin{center}
\begin{figure}[H]
\includegraphics[scale=.8731]{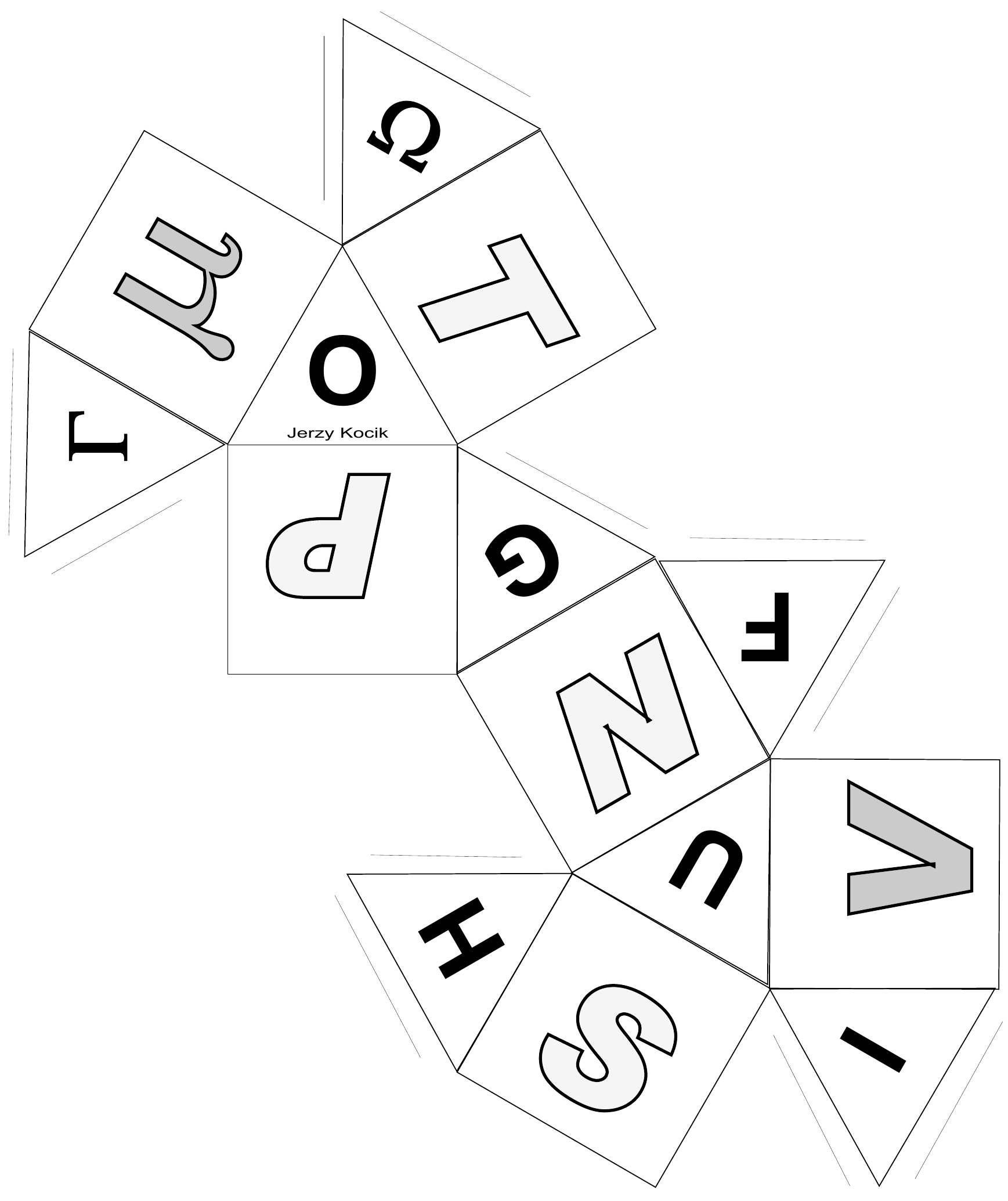}
\caption{Magic cube of thermodynamics -- enlarge, cut and glue.  
For consistency with the text, ``$P$'' should be replaced by ``$\mP \equiv -P$''}
\end{figure}
\end{center}

\subsection*{TABLE 4.  Gibbs' relations:}

$$
\begin{aligned}
T  &\ = \ {{\q U} \over \q S} \midx_{N,V}
            = {{\q H} \over \q S} \midx_{N,P}
                       =  {{\q I} \over \q S} \midx_{\mu,V}
                              =  {{\q \Gamma} \over \q S} \midx_{P,\mu}\\
-P  &\ = \  {{\q U} \over \q V} \midx_{S,N}
            = {{\q F} \over \q V} \midx_{T,N}
                       = {{\q \Omega} \over \q V} \midx_{T,\mu}
                              = {{\q I} \over \q V} \midx_{S,\mu}\\
\mu&\ = \  {{\q U} \over \q N} \midx_{V,S}
            = {{\q H} \over \q N} \midx_{S,P}
                       =  {{\q G} \over \q N} \midx_{T,P}
                              =  {{\q F} \over \q N} \midx_{V,T}\\
S  &\ = \  -{{\q F} \over \q T} \midx_{V,N}
            = -{{\q G} \over \q T} \midx_{P,N}
                              = -{{\q \Omega} \over \q T} \midx_{V,\mu}\\
V  &\ = \  {{\q H} \over \q P} \midx_{N,S}
            = {{\q G} \over \q P} \midx_{T,N}
                       =  {{\q \Gamma} \over \q P} \midx_{S,\mu}\\
N  &\ = \  -{{\q \Omega} \over \q \mu} \midx_{V,T}
            = -{{\q \Gamma} \over \q \mu} \midx_{S,P}
                       =  -{{\q I} \over \q \mu} \midx_{S,V} 
\end{aligned}
$$

\subsection*{TABLE 5. Gibbs differential form:}

$$
\begin{aligned}
dU      & \ = \  T\,dS + \mP\,dV + \mu\,dN  \\
dF      &\ = \  -S\,dT + \mP\,dV + \mu\,dN  \\
dI      &\ = \  T\,dS + \mP\,dV - N\,d\mu  \\
dH      & \ = \  T\,dS - V\,d\mP + \mu\,dN \\
d\Gamma & \ = \ T\,dS - V\,d\mP - N\,d\mu  \\
dG      & \ = \  -S\,dT - V\,d\mP + \mu\,dN  \\
d\Omega & \ = \  -S\,dT + \mP\,dV - N\,d\mu  \\
0       & \ = \ -S\,dT - V\,d\mP - N\,d\mu  
\end{aligned}
$$

\subsection*{TABLE 6.  Maxwell's identities:}

$$
\begin{aligned}
{\q V \over\q T } &  \ = \  {\q S \over\q \mP  } \midx_{T,P,\;(\mu\hbox{~or~} N)} \\
{\q V \over\q\mu} & \ = \  {\q N \over\q \mP  } \midx_{\mu,P,\;(T \hbox{~or~} S)} \\
 {\q N \over\q T } & \ = \  {\q S \over\q \mu} \midx_{T,\mu,\;(P \hbox{~or~} V)} \\
 -{\q \mP \over\q T } & \ = \  {\q S \over\q V  } \midx_{T,V, \; (\mu\hbox{~or~} N)} \\
- {\q N \over\q V } & \ = \  {\q \mP \over\q \mu} \midx_{V,\mu,\;(T \hbox{~or~} S)} \\
-{\q T \over\q\mu} & \ = \  {\q N \over\q S  } \midx_{\mu,S,\;(V \hbox{~or~} P)} 
\end{aligned}
\qquad\qquad
\begin{aligned}
{\q T \over\q V } & \ = \  {\q \mP \over\q S } \midx_{V,S,\;(\mu \hbox{~or~} N)} \\
{\q\mu\over\q V } & \ = \  {\q \mP \over\q N } \midx_{V,N,\;(T  \hbox{~or~} S)} \\
 {\q T \over\q N } & \ = \  {\q\mu\over\q S } \midx_{N,S,\;(P \hbox{~or~} V)} \\
- {\q T \over\q \mP } & \ = \  {\q V \over\q S } \midx_{P,S,\;(\mu \hbox{~or~} N)} \\
- {\q V \over\q N } & \ = \  {\q\mu\over\q \mP } \midx_{N,P,\; (S \hbox{~or~} T)} \\
-{\q\mu\over\q T } & \ = \  {\q S \over\q N } \midx_{T,N,\;(V \hbox{~or~} P)} 
\end{aligned}
$$

%
%
%


%
%
%


\vfill

\hrule 

~

\noindent
{\bf Note:} 
This text is almost an exact copy of the paper that appeared 
in "Symmetries in Science II," (B.Gruber, ed.), Plenum, New York, 1986, pp 279-287, 
except a part of the first chapter was moved to the end for clarity, and a short
appendix on "thermodynamical cube" from the first submission preserved here
(presented as a poster ``Magic Cube of Thermodynamics'' at the Symposium on Symmetries in Science II,  
Carbondale, Illinois, March 1986.

\end{document}